\documentclass[aps,prl,preprint,tightenlines,superscriptaddress,showpacs,byrevtex]{revtex4}

\usepackage{graphicx}
\usepackage{color}

\begin{document}


\preprint{\vbox{
                 \hbox{} 
}}

\title{ \quad\\[0.5cm] Search for Lepton and Baryon Number Violating $\tau^-$ Decays into $\bar{\Lambda}\pi^-$ and $\Lambda \pi^-$}

\affiliation{Budker Institute of Nuclear Physics, Novosibirsk}
\affiliation{Chiba University, Chiba}
\affiliation{Chonnam National University, Kwangju}
\affiliation{University of Cincinnati, Cincinnati, Ohio 45221}
\affiliation{University of Hawaii, Honolulu, Hawaii 96822}
\affiliation{High Energy Accelerator Research Organization (KEK), Tsukuba}
\affiliation{Hiroshima Institute of Technology, Hiroshima}
\affiliation{Institute of High Energy Physics, Chinese Academy of Sciences, Beijing}
\affiliation{Institute of High Energy Physics, Vienna}
\affiliation{Institute for Theoretical and Experimental Physics, Moscow}
\affiliation{J. Stefan Institute, Ljubljana}
\affiliation{Kanagawa University, Yokohama}
\affiliation{Korea University, Seoul}
\affiliation{Kyungpook National University, Taegu}
\affiliation{Swiss Federal Institute of Technology of Lausanne, EPFL, Lausanne}
\affiliation{University of Ljubljana, Ljubljana}
\affiliation{University of Maribor, Maribor}
\affiliation{University of Melbourne, Victoria}
\affiliation{Nagoya University, Nagoya}
\affiliation{Nara Women's University, Nara}
\affiliation{National Central University, Chung-li}
\affiliation{National United University, Miao Li}
\affiliation{Department of Physics, National Taiwan University, Taipei}
\affiliation{H. Niewodniczanski Institute of Nuclear Physics, Krakow}
\affiliation{Nippon Dental University, Niigata}
\affiliation{Niigata University, Niigata}
\affiliation{Nova Gorica Polytechnic, Nova Gorica}
\affiliation{Osaka City University, Osaka}
\affiliation{Osaka University, Osaka}
\affiliation{Panjab University, Chandigarh}
\affiliation{Peking University, Beijing}
\affiliation{Princeton University, Princeton, New Jersey 08544}
\affiliation{University of Science and Technology of China, Hefei}
\affiliation{Seoul National University, Seoul}
\affiliation{Shinshu University, Nagano}
\affiliation{Sungkyunkwan University, Suwon}
\affiliation{University of Sydney, Sydney NSW}
\affiliation{Tata Institute of Fundamental Research, Bombay}
\affiliation{Toho University, Funabashi}
\affiliation{Tohoku Gakuin University, Tagajo}
\affiliation{Tohoku University, Sendai}
\affiliation{Department of Physics, University of Tokyo, Tokyo}
\affiliation{Tokyo Institute of Technology, Tokyo}
\affiliation{Tokyo Metropolitan University, Tokyo}
\affiliation{Tokyo University of Agriculture and Technology, Tokyo}
\affiliation{University of Tsukuba, Tsukuba}
\affiliation{Virginia Polytechnic Institute and State University, Blacksburg, Virginia 24061}
\affiliation{Yonsei University, Seoul}

  \author{Y.~Miyazaki}\affiliation{Nagoya University, Nagoya} 
  \author{K.~Abe}\affiliation{High Energy Accelerator Research Organization (KEK), Tsukuba} 
 \author{K.~Abe}\affiliation{Tohoku Gakuin University, Tagajo} 
  \author{I.~Adachi}\affiliation{High Energy Accelerator Research Organization (KEK), Tsukuba} 
  \author{H.~Aihara}\affiliation{Department of Physics, University of Tokyo, Tokyo} 
  \author{K.~Arinstein}\affiliation{Budker Institute of Nuclear Physics, Novosibirsk} 
  \author{Y.~Asano}\affiliation{University of Tsukuba, Tsukuba} 
  \author{V.~Aulchenko}\affiliation{Budker Institute of Nuclear Physics, Novosibirsk} 
  \author{T.~Aushev}\affiliation{Institute for Theoretical and Experimental Physics, Moscow} 

  \author{S.~Bahinipati}\affiliation{University of Cincinnati, Cincinnati, Ohio 45221} 
% ...
\collaboration{The Belle Collaboration}

\begin{abstract}
We have searched for 
$\tau$ lepton decays,
$\tau^-\rightarrow \bar{\Lambda}\pi^-$ and $\tau^-\rightarrow \Lambda \pi^-$,
{that simultaneously 
violate the conservation of 
the lepton ($L$) and baryon ($B$) number 
using a data sample of
154 fb$^{-1}$ collected with
the Belle detector at the KEKB $e^+e^-$ asymmetric-energy collider.}
No evidence for a signal {was} found
in either of the decay modes
and we set the following upper limits for the branching fractions:
${\cal{B}}(\tau^-\rightarrow \bar{\Lambda}\pi^-) < 1.4\times 10^{-7}$
and 
${\cal{B}}(\tau^-\rightarrow \Lambda \pi^-) < 0.72\times 10^{-7}$ 
at the 90\% confidence level. 
This is the first search  ever performed for these modes.
\end{abstract}


\maketitle

\section{Introduction}

While the Standard Model (SM) assumes
both the baryon ($B$) and lepton number ($L$) conservation, 
in some extensions of the SM, such as 
supersymmetric~\cite{cite:susy} and 
superstring~\cite{cite:string} models,
$B$ and $L$ violation is expected 
while their difference $B-L$ should be conserved.
Existing searches for baryon number violation are restricted to
experiments with nucleons~\cite{PDG}.
Limits from searches for proton decay imply that
the $\tau$ lepton decays with baryon number violation 
can be expected to have a probability well below 
the current experimental sensitivity~\cite{cite:tau02}.
However,
high luminosity $B$-factories provide 
an opportunity to look for decays of the $\tau$ lepton,
$D$ and $B$ mesons that violate $B$ and $L$ 
with unprecedented sensitivity.
Until now,
searches for $\tau$ lepton decays with a baryon final 
state involved a proton only~\cite{protontau}.
Upper limits for the branching fraction of these decays
are in the range of $10^{-5} \sim 10^{-6}$. 
In a recent extension of the SM~\cite{cite:Hou:2004uc},
decays of $\tau$ lepton as well as $D$ and $B$ mesons
that violate $B$ and $L$ were considered 
using
right-handed four-fermion couplings that conserve 
$B-L$, 
these suggested that $\tau^-\rightarrow \bar{\Lambda}\pi^-$ 
decay involving
the second and third generation transition might be interesting.

We report here our {search} for
$\tau^-\rightarrow\bar{\Lambda}\pi^-$ and $\tau^-\rightarrow\Lambda\pi^-$ decays
with a data sample
of 154 fb$^{-1}$ 
collected at the $\Upsilon(4S)$ resonance
and 60 MeV below it
with the Belle detector at the KEKB  $e^+e^-$ 
asymmetric-energy collider~\cite{kekb}. 
These $\tau$ lepton decay modes 
have never been studied before
\footnotemark[2]
\footnotetext[2]{Unless otherwise stated, charge 
conjugate decays are implied throughout
this paper.}.

The Belle detector is a large-solid-angle magnetic spectrometer that
consists of a silicon vertex detector (SVD), 
a 50-layer central drift chamber (CDC), 
an array of aerogel threshold \v{C}erenkov counters (ACC), a barrel-like arrangement of 
time-of-flight scintillation counters (TOF), and an electromagnetic calorimeter 
comprised of CsI(Tl) crystals (ECL) located inside
a superconducting solenoid coil
that provides a 1.5~T magnetic field.  
An iron flux-return located outside of the coil is instrumented to detect $K_L^0$ mesons 
and to identify muons (KLM).  
The detector is described in detail elsewhere~\cite{Belle}.

Particle identification 
is very important in this measurement.
It is based on 
the ratio of the energy 
deposited in the ECL to the momentum measured in the SVD and CDC, 
the shower shape in the ECL, 
the particle range in the KLM, 
the hit information from the ACC,
$dE/dX$ in the CDC 
and time-of-flight from the TOF.
To distinguish hadron species,
we use likelihood ratios,
for instance, 
${\cal{P}}(p/\pi) = {\cal{L}}_p/({\cal{L}}_p + {\cal{L}}_{\pi} )$,
where ${\cal{L}}_i$ is the likelihood for the detector response 
to the track with flavor hypothesis $i$. 
For lepton identification,
we use  likelihood ratios as electron probability ${\cal P}(e)$ \cite{EID} and 
muon probability ${\cal P}({\mu})$ \cite{MUID} 
determined by 
{the responses of the appropriate subdetectors.}

For the Monte Carlo (MC) studies,
the following programs are used to
generate background events:
KORALB/TAUOLA~\cite{cite:koralb_tauola} for $\tau^+\tau^-$, 
QQ~\cite{cite:qq} for $B\bar{B}$ and continuum,
BHLUMI~\cite{BHLUMI} for Bhabha,
KKMC~\cite{KKMC} for $\mu^+\mu^-$ and
AAFH~\cite{AAFH} for two-photon processes.
Signal MC is generated by KORALB/TAUOLA.
To take a range of possible $\tau^-\rightarrow \bar{\Lambda}\pi^-$
and $\tau^-\rightarrow \Lambda \pi^-$ decay mechanisms into account,
we generate events under three different assumptions:
uniform angular distribution in the $\tau$ rest frame,
$V-A$ and $V+A$ interactions.
We initially assume all $\tau$ decays to have a uniform angular distribution
and take the other hypotheses into account later.
The Belle detector response is simulated by a GEANT 3~\cite{cite:geant3} 
based program.
Unless stated otherwise,
all kinematic variables are
calculated in the laboratory frame,
while the ones calculated in the $e^+e^-$ center-of-mass (CM) frame 
are indicated by the superscript ``CM''.


%
%

\section{Data Analysis}

We search for $\tau^+\tau^-$ events
in which one $\tau$ decays
into $\bar{\Lambda}\pi^-$ and ${\Lambda}\pi^-$ (signal side) 
and the other $\tau$ (tag side) decays into one charged particle 
with any number of
additional photons and neutrinos. 
Thus, for the $B-L$ conserving modes the experimental signature is
\begin{center}
$\left\{\tau^- \rightarrow (\bar{p}\pi^+) + \pi^- \right\} ~+
 ~ \left\{ \tau^+ \rightarrow ({\rm a~track})^+ + (n^{\rm TAG}_{\gamma} \ge 0)
 + X(\rm{missing}) \right\}$,
\end{center}
and for the $B-L$ violating modes,
it is
\begin{center}
$\left\{\tau^- \rightarrow (p\pi^-) + \pi^- \right\} ~+~ 
 \left\{ \tau^+ \rightarrow ({\rm a~track})^+ + (n^{\rm TAG}_{\gamma} {\ge} 0)
 + X(\rm{missing}) \right\}$.
\end{center}
Here charged tracks and photons 
are {required} to be reconstructed within the fiducial volume 
defined by $-0.866 < \cos\theta < 0.956$,
where $\theta$ is the polar angle with
respect to the direction opposite to the $e^+$ beam.
Charged tracks should have
momentum transverse to the $e^+$ beam
$p_t > 0.1$ GeV/$c$ 
and 
photons should have energies
$E_{\gamma} > 0.1$ GeV.
We denote the pion from $\tau\rightarrow\Lambda\pi$ as $\pi_1$ and 
that from $\Lambda\rightarrow p\pi$ as $\pi_2$. 
We can distinguish between the $B-L$ conserving and violating modes
by the charge of these pions:
the $B-L$ conserving decay modes have an opposite sign combination
on the $\pi_1$ and $\pi_2$ charges, 
while the $B-L$ violating modes have a same sign combination.

\begin{figure}[h]
\begin{center}
 \resizebox{0.45\textwidth}{0.35\textwidth}{\includegraphics
 {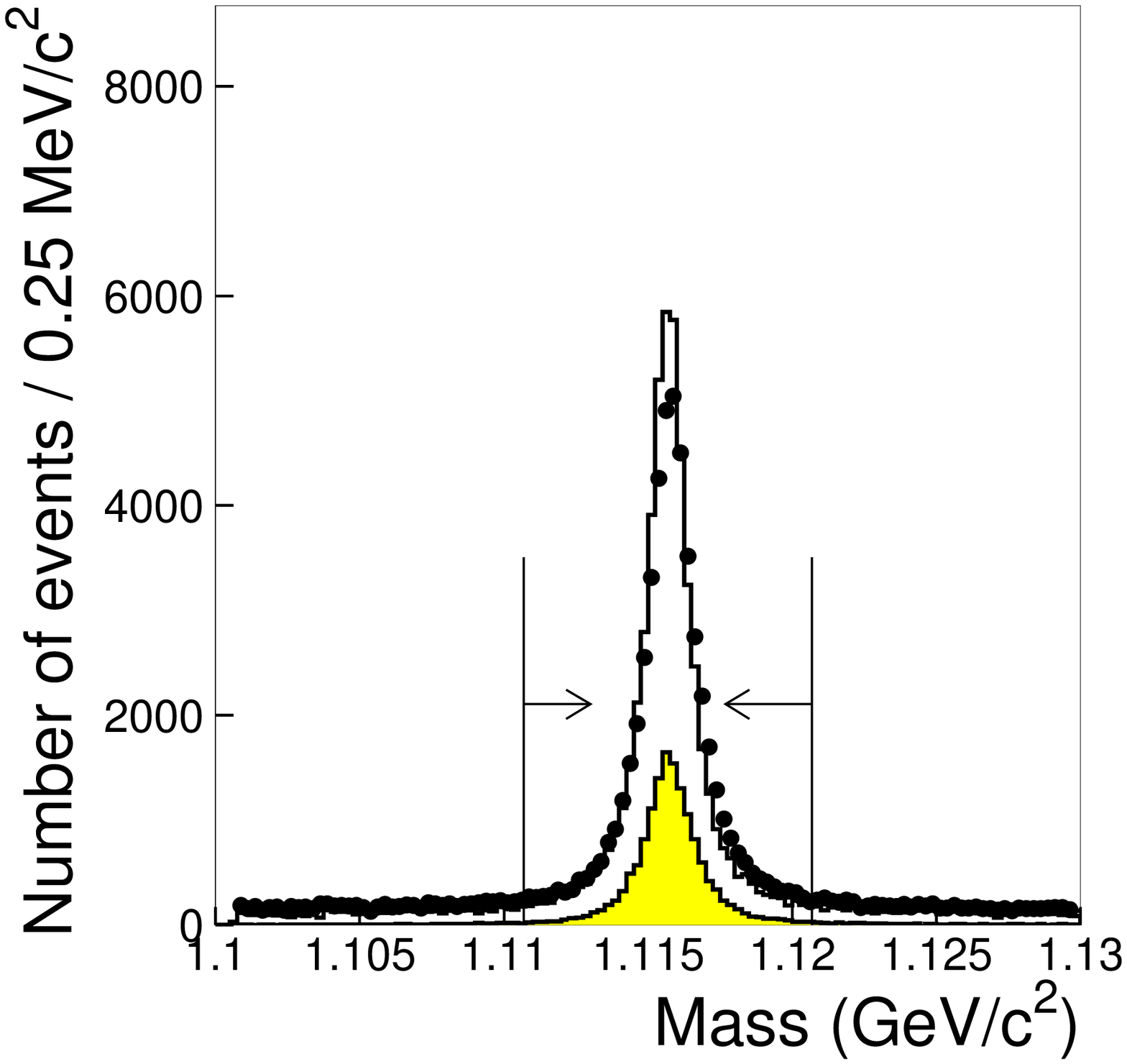}}
 \resizebox{0.45\textwidth}{0.35\textwidth}{\includegraphics
 {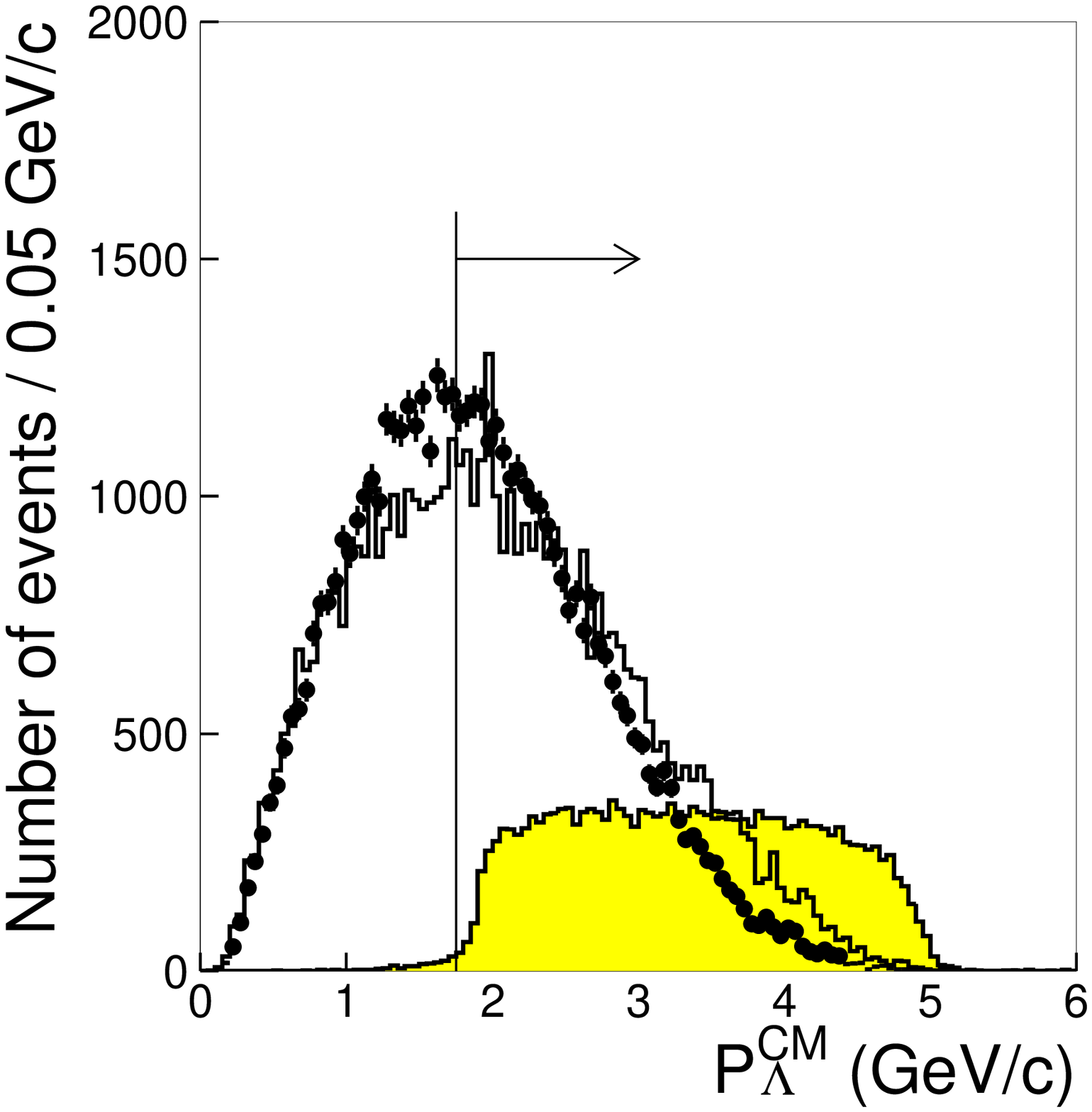}}
 \caption{ Reconstructed ${\Lambda}$ candidate mass (left) 
 and momentum (right) in the CM frame.
 The signal MC distributions are indicated by the filled histogram, 
 all background including $\tau^+\tau^-$ and $q\bar{q}$ by the open histogram, 
 and closed circles are data.
 While the signal MC distribution is normalized arbitrarily, 
 the data and MC are normalized to the same luminosity.
 The selected area is indicated by the vertical lines with arrows. 
}
\label{fig:lambdainfo}
\end{center}
\end{figure}

We first demand that the four tracks have a zero net charge.  The
magnitude of the thrust~\cite{thrust} is required to be larger than 0.9 to suppress
the $q\bar{q}$ continuum background.  The event should have a 1-3
prong configuration relative to the plane perpendicular to the thrust
axis.  
We select $\Lambda$ candidates via the
${p}\pi^-$ decay channel based on the angular difference
between the ${\Lambda}$ flight direction and the direction
pointing from the interaction point to the decay vertex (see
Ref.\cite{Lambda_rec} for more details).
The proton from the ${\Lambda}$ decay is identified by 
a condition
${\cal{P}}(p/\pi) > 0.6$, 
{which has a 70\% efficiency
and
11\% and 4\% fake rate from $K$ and $\pi$, respectively.}
In order to avoid fake ${\Lambda}$ candidates in which 
an $e^+e^-$ pair from a photon conversion
{results in}
two tracks
in the signal side,
an electron veto (${\cal P}(e) < 0.1$) is imposed on the three tracks on the signal side.
The reconstructed ${\Lambda}$ candidate mass should be
within $\pm5$ MeV/$c$$^2$ of the nominal $\Lambda$ mass and
its momentum,
$p^{\rm{CM}}_{\Lambda}$, 
is required
to be larger than 1.75 GeV/$c$
to reduce contributions from the generic $\tau^+\tau^-$ and $q\bar{q}$ continuum 
background
as shown in Fig.~\ref{fig:lambdainfo}.       
A discrepancy between the data and MC observed in the $\Lambda$ momentum
distribution in the CM system shown in Fig.~\ref{fig:lambdainfo} (right)
is apparently caused by the imperfect simulation of baryon spectra
in the QQ generator.
Since the final estimate of the background uses
information from the sideband data,
this discrepancy does not directly affect our results.

\begin{figure}[h]
\begin{center}
 \resizebox{0.45\textwidth}{0.35\textwidth}{\includegraphics
 {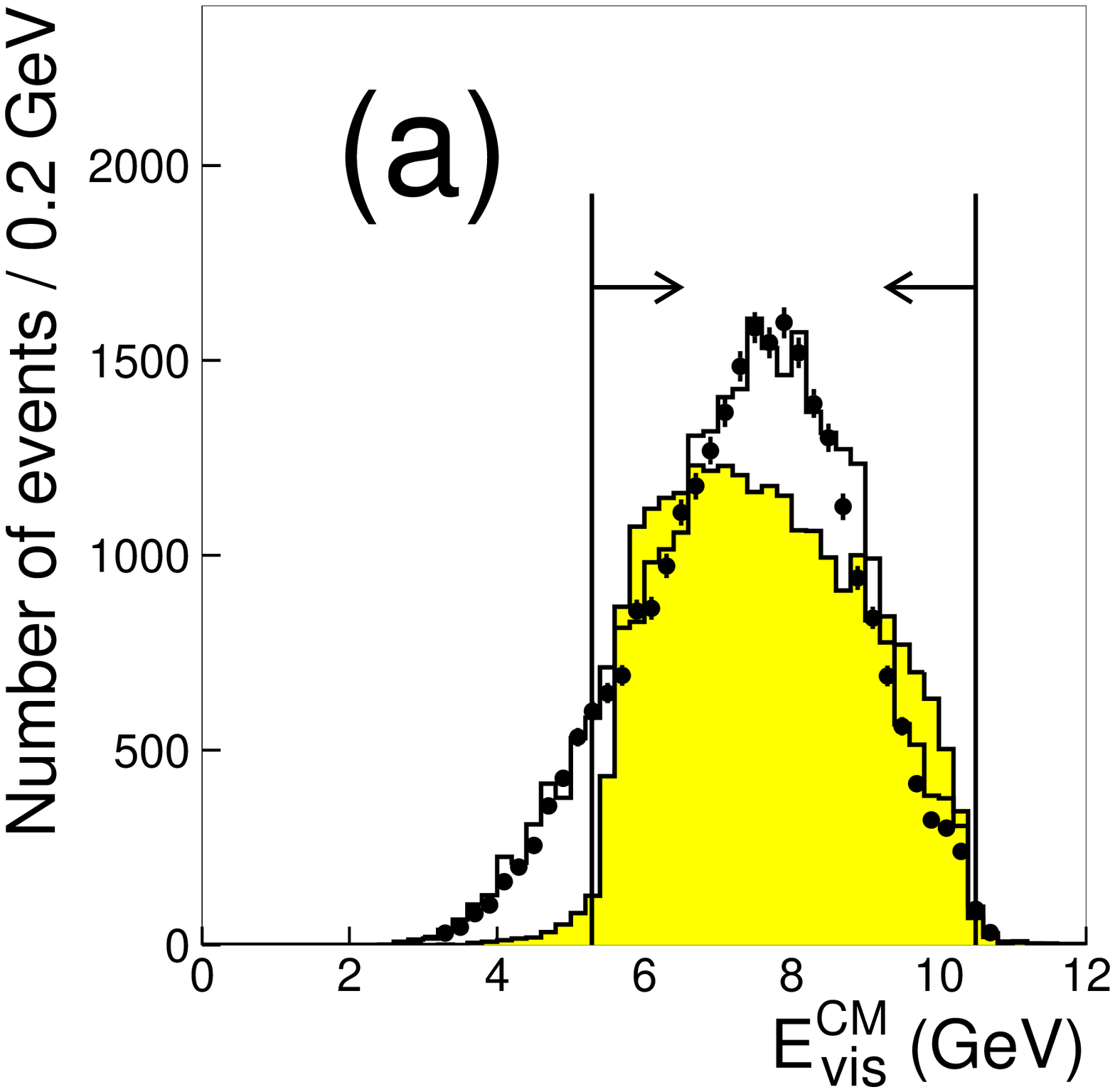}}
 \resizebox{0.45\textwidth}{0.35\textwidth}{\includegraphics
 {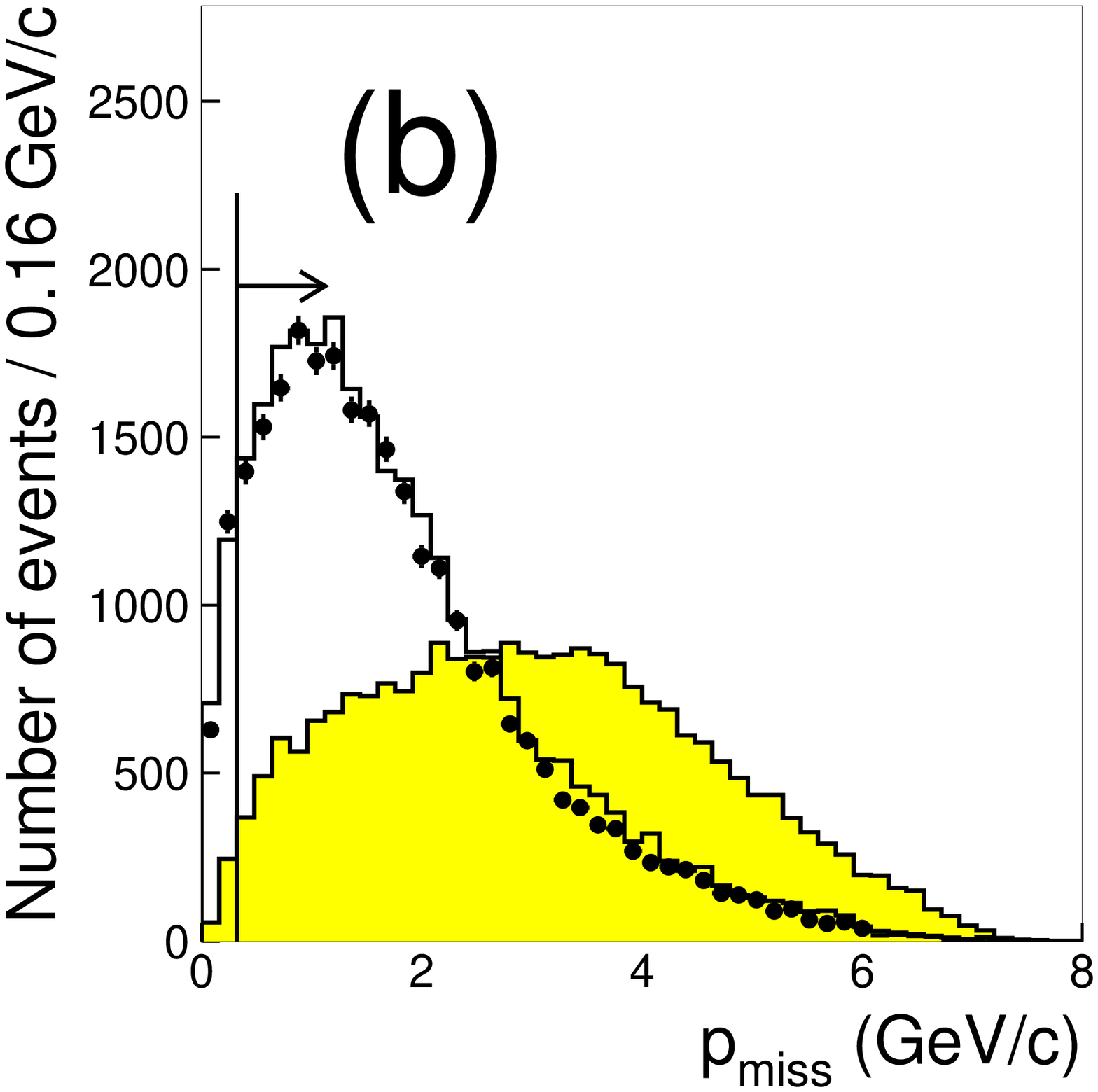}}
 \resizebox{0.45\textwidth}{0.35\textwidth}{\includegraphics
 {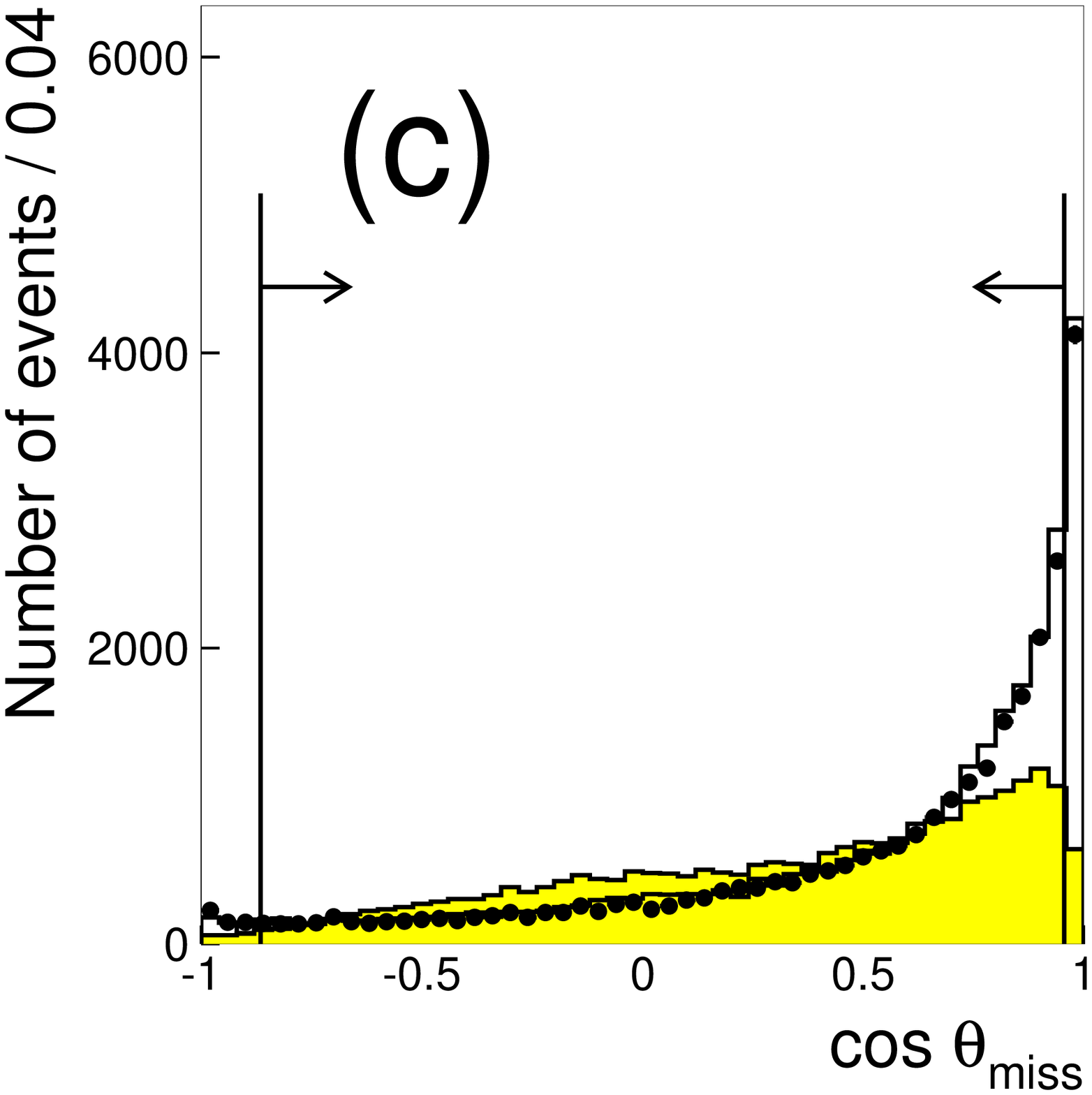}}
 \resizebox{0.45\textwidth}{0.35\textwidth}{\includegraphics
 {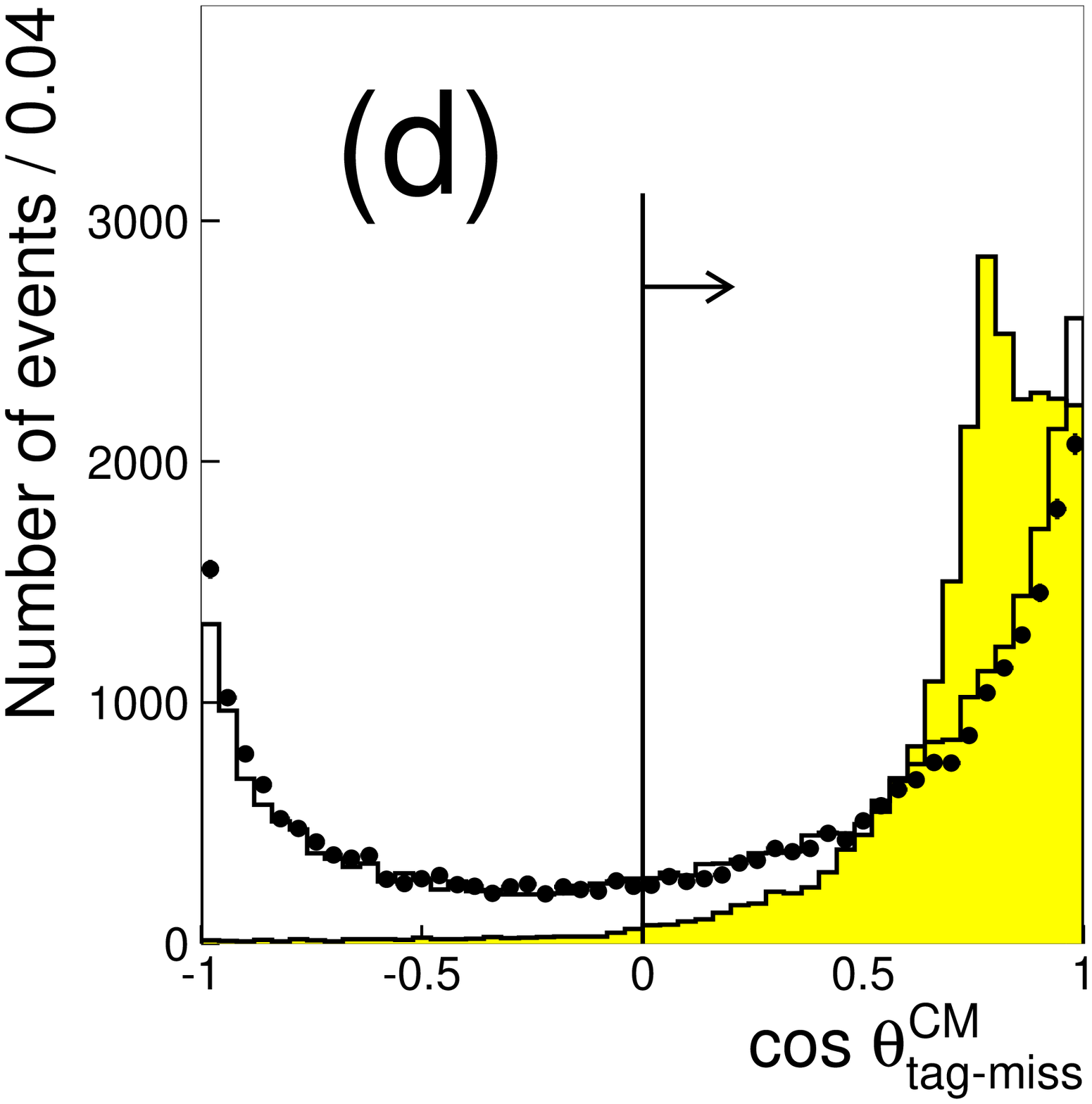}}
 \caption{ 
 Kinematical distributions used in the event selection
 after $\Lambda$ mass and momentum selection:
 (a) the total visible energy in the CM frame; 
 (b) the momentum of the missing particle; 
 (c) the polar angle of the missing particle;
 (d) the opening angle between the missing particle and
 tag-side track in the CM frame.
 The signal MC distributions are indicated by the filled histograms, 
 the total background including $\tau^+\tau^-$ and $q\bar{q}$ is shown by 
 the open histogram, 
 and closed circles are data.
 While the signal MC distribution is normalized arbitrarily, 
 the data and MC are normalized to the same luminosity.
 The selected area is indicated by the vertical lines with arrows. 
}
\label{fig:cut}
\end{center}
\end{figure}

The total visible energy in the CM frame  $E^{\rm{CM}}_{\rm{vis}}$,
defined as the sum of the energies
of the $\Lambda$ candidate,
two other tracks 
and photon candidates, 
is required to satisfy
$5.29$ GeV $< E^{\rm{CM}}_{\rm{vis}} < 10.5$ GeV 
(see Fig.~\ref{fig:cut} (a)).
To ensure that the missing particles
are only neutrino(s) {
rather than photons} or charged particles that fall outside
the detector acceptance,
we require
the residual momentum vector  $\vec{p}_{\rm miss}$  
to have a magnitude above  0.4 GeV/$c$ 
and 
to point into the fiducial volume of 
the detector:
$-0.866 < \cos\theta_{\rm{miss}} < 0.956$
(see Fig.~\ref{fig:cut} (b) and (c)).
We select  events with
missing particles
belonging to the tag side 
by requiring 
the opening angle between the residual momentum vector and
tag-side track to be smaller than 90 degrees 
{in the CM system}
(see Fig.~\ref{fig:cut} (d)).
To suppress the continuum background,
the following requirements are imposed: 
the number of the photon candidates on the signal and tag side,
$n_{\gamma}^{\rm{SIG}}\leq 1$ and $n_{\gamma}^{\rm{TAG}}\leq 2$,
respectively.
Both the proton veto  
(${\cal P}({p}/\pi) < 0.6$)
and kaon veto
(${\cal P}(K/\pi) < 0.6$)
are applied to $\pi_1$
and the tag-side track.

\begin{figure}[t]
\begin{center}
 \resizebox{.8\textwidth}{!}{\includegraphics
 {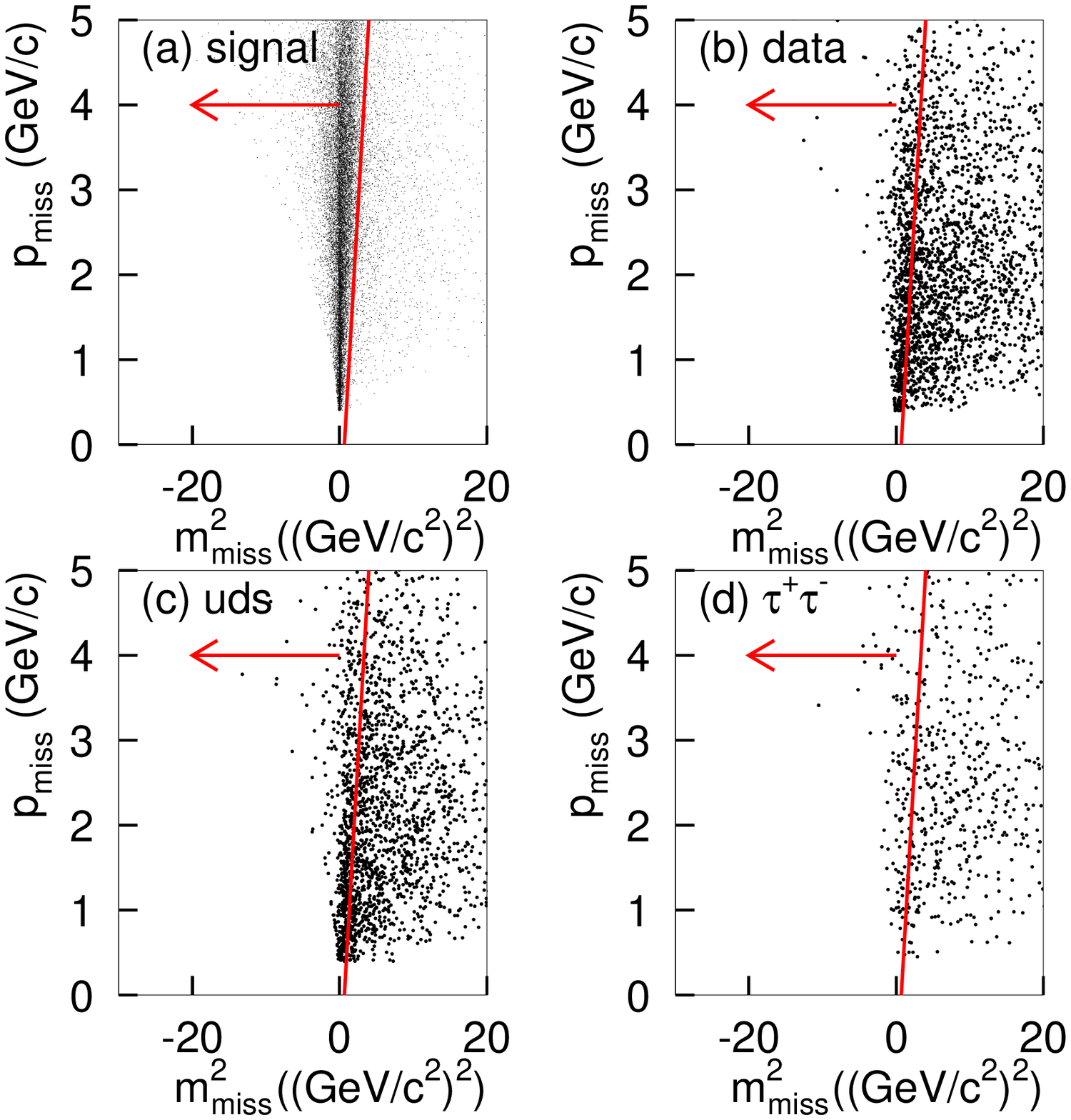}}
 \caption{$p_{\rm{miss}}$ vs $m_{\rm{miss}}^2$ plots
 for (a) signal MC events,  (b) data,  
(c) $q\bar{q}$ ($uds$) continuum MC events 
 and $\tau^+\tau^-$ MC events. 
The area indicated by the arrow is the selected region. 
 \label{fig:pmiss_vs_mmiss2}
 }
 \end{center}
\end{figure}

Finally, a condition is imposed on
the relation between
$p_{\rm{miss}}$ and 
{the mass squared} of
the missing particle, $m^2_{\rm miss}$.
The latter is defined as 
$E^2_{\rm miss}-p^2_{\rm miss}$,
where $E_{\rm miss} = E_{\rm total}-E_{\rm vis}$, 
$E_{\rm total}$ is the sum of the beam energies (11.5 GeV) 
and $E_{\rm vis}$ is 
the sum of all visible energy.
To further suppress background from generic $\tau^+\tau^-$
and 
continuum background,
we require
the following relation
between the missing momentum $p_{\rm{miss}}$ and 
$m^2_{\rm{miss}}$ :
$p_{\rm{miss}} > 1.5[\mbox{GeV}^{-1}c^3]\times m^2_{\rm{miss}}- 1.0[\mbox{GeV}/c]$
(see Fig. \ref{fig:pmiss_vs_mmiss2}). 
While this condition retains 89\% of the signal events,
81\% of the generic $\tau^+\tau^-$ and 77\% of $uds$ continuum background
are removed.  
There is no significant difference between signal distributions 
for the $B-L$ conserving and violation modes according to the MC. 
After applying the above  requirements,
{82 and 62 events are retained
in the data for the $B-L$ conserving 
and violating mode, respectively.}

\section{Results}

Signal candidates are examined in the two-dimensional 
space of the $\bar{\Lambda}\pi^-$/ $\Lambda\pi^-$ invariant 
mass, $M_{\rm {inv}}$, and the difference of their energy from the 
beam energy in the CM system, $\Delta E$.
A signal event should have $M_{\rm {inv}}$
close to the $\tau$-lepton mass
and 
$\Delta E$ close to 0.
The  $M_{\rm {inv}}$ and  $\Delta E$  resolutions are parameterized 
from the MC distributions around the peak using an asymmetric
Gaussian shape to account for initial state radiation 
with widths 
$\sigma^{\rm{high/ low}}_{M_{\rm{inv}}} = 4.6/ 4.0$ MeV/$c$$^2$ and 
$\sigma^{\rm{high/ low}}_{\Delta E} = 22/ 29$ MeV,
where the ``high/low'' superscript indicates the higher/lower side 
of the peak.

\begin{figure}[t]
\begin{center}
 \resizebox{0.45\textwidth}{0.45\textwidth}{\includegraphics
 {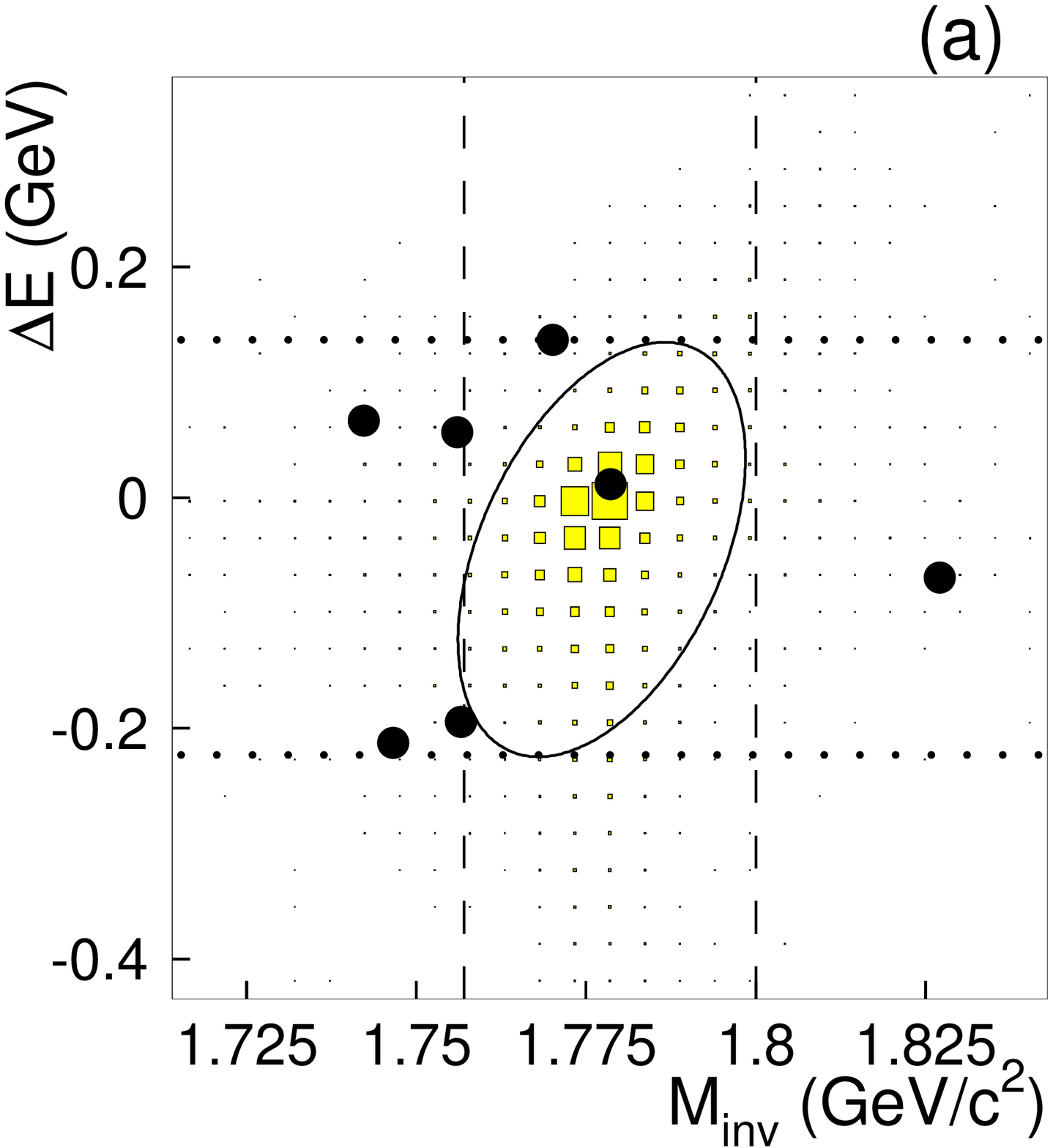}}
 \resizebox{0.45\textwidth}{0.45\textwidth}{\includegraphics
 {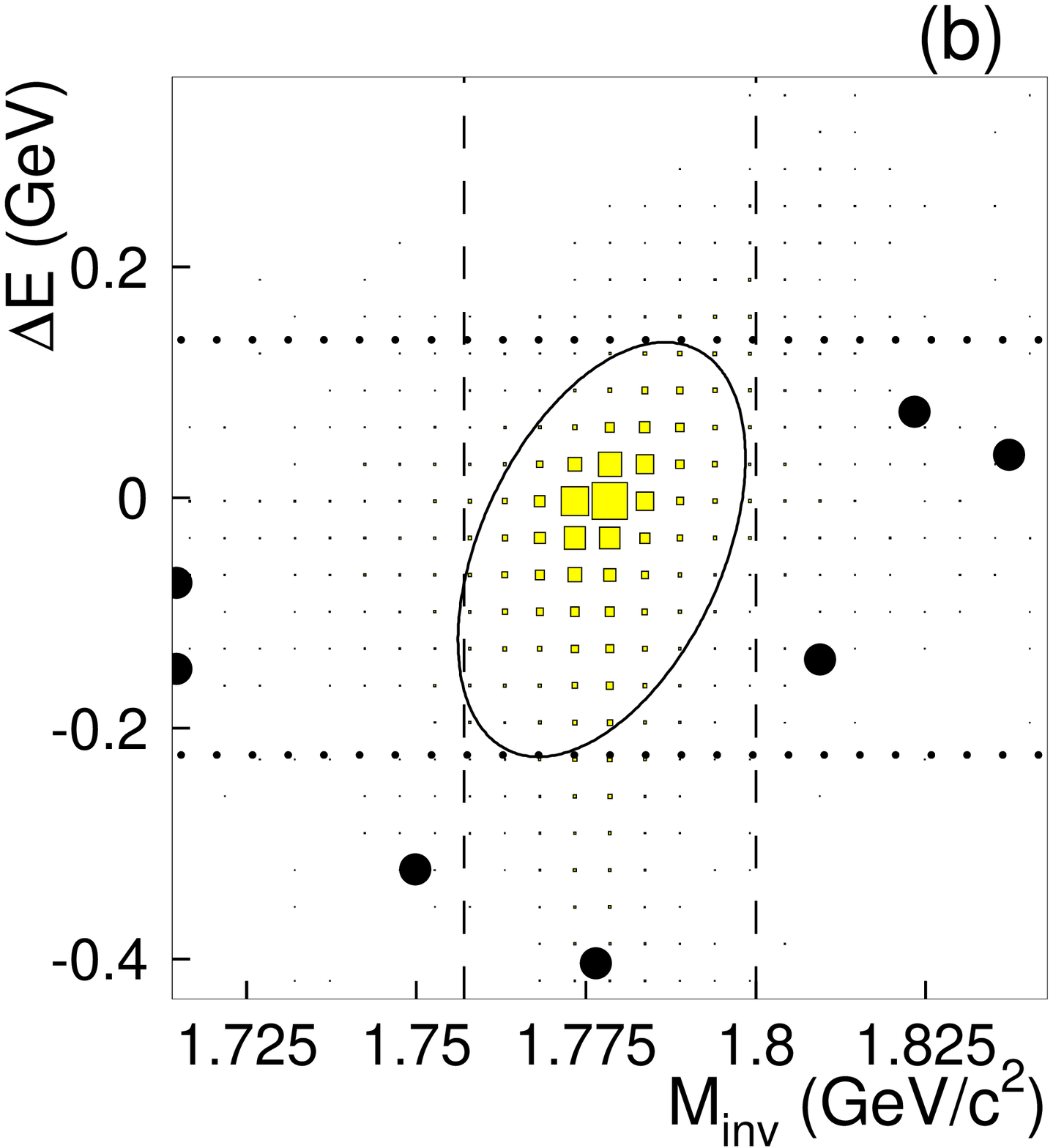}}
 \caption{
Scatter-plot of data in the $M_{\rm inv}$ -- $\Delta{E}$
plane: (a) and (b) correspond to
the $\pm 15 \sigma$ area for
the $B-L$ conserving and violating modes, respectively.
The 90\% elliptical region shown by a solid curve in (a) and (b) 
is used for evaluating the signal yield.
In (a) and (b), the vertical dashed lines denote 
the boundaries of
the blind regions,
while the regions inside the horizontal dotted lines and outside the vertical dashed lines
are sidebands 
used to estimate the expected background in the elliptical region.
Closed circles correspond to the data.
The filled boxes show the MC signal distribution
with arbitrary normalization.
\label{fig:5}
}
\end{center}
\end{figure}

We blind a region of $\pm 5\sigma_{\rm{M_{inv}}}$ 
around the $\tau$ mass in $\rm{M_{\rm inv}}$ 
and 
a region of
$-0.5 < \Delta E < 0.5$ GeV
so as not to bias our choice of selection criteria.
{Figure~\ref{fig:5}} shows scatter-plots 
for data and MC samples 
distributed over $\pm 15\sigma$ 
in the $M_{\rm{inv}}-\Delta E$ plane.
For the $B-L$ conserving mode, 
{the number of data events and 
the total number of events expected from 
MC}  
outside the blinded region
(bounded by the vertical dotted line in Fig.~\ref{fig:5} (a) and (b))
is $5$ and $8.5 \pm 3.1$ events, respectively,
while 
{that} 
for the  $B-L$ violating mode 
{is} $6$ and $5.1 \pm 2.3$ events, respectively,
indicating good agreement between the data and 
MC describing the background.
The surviving
background events are due to generic $\tau^+\tau^-$ decays (about 1/2) and
$uds$ continuum
(about 1/2). 
The former events are dominated by
the $\tau \to a_1(1260) \nu_{\tau}$ decays, in which 
two of the
three
charged pions from the $a_1(1260)$ decay form a fake $\Lambda$ candidate.
The continuum background events have one true ${\Lambda}$ that forms a
signal candidate together with another track.

As a signal region, 
we take an elliptically shaped area that contains 90\% of MC
signal events remaining 
after all the  requirements
in $M_{\rm inv}-\Delta E$ plane, 
as shown in Fig.~\ref{fig:5} (a) and (b).
It corresponds to
a signal detection efficiency of $11.8\pm0.1\%$
for the $B-L$ conserving and $11.7\pm0.1\%$ for the $B-L$ violating modes.

We assume the background distribution to be flat along the
$M_{\rm{inv}}$ axis, and then obtain the expected background in the ellipse 
to be $1.7 \pm 0.8$ events for each of the two modes,
using sideband regions 
inside the horizontal dotted lines and outside the vertical dotted lines,
as shown in  Fig.~\ref{fig:5}  (a) and (b).
We open the blinded region and find
only one data event in the ellipse for the $B-L$ conserving mode
and 
no data events for the $B-L$ violating mode 
(see Fig.~\ref{fig:5} (a) and (b), respectively). 
Since no statistically significant excess of data over
the expected background in the signal region is observed,
we apply the frequentist approach 
to calculate {an} upper limit for the signal yield~\cite{cite:FC}.
The resulting limits for 
the signal yields at 90\% confidence level, $s_{90}$,  
are 
$2.76$ 
and  
$1.30$, 
respectively. 
The upper limits on the branching fraction
before the inclusion of systematic uncertainties are then
calculated as
\begin{equation}
{\cal B}(\tau \rightarrow \Lambda \pi) 
<  \frac{s_{90}}{2 \varepsilon N_{\tau\tau}{\cal B}
( \Lambda \rightarrow p \pi^-)}
\end{equation}
where $N_{\tau\tau} = 137 \times 10^6$ and 
${\cal B}(\Lambda \rightarrow p \pi^-) = 0.639$~\cite{PDG}. 
The resulting values are
${\cal B}(\tau^-\rightarrow\bar{\Lambda}\pi^-) < 1.3\times 10^{-7}$
and 
${\cal B}(\tau^-\rightarrow \Lambda\pi^-) < 0.64 \times 10^{-7}$. 

%
%

The sources of
systematic uncertainty in the detection sensitivity 
$2\varepsilon N_{\tau\tau}{\cal B}(\Lambda\rightarrow p\pi^-)$ 
are listed in Table \ref{tbl:sys}.
{The} uncertainty from $\Lambda$ reconstruction is
estimated to be 6.0\% 
from the difference of the 
proper time distributions 
for data and MC simulation.
The proton identification 
in $\Lambda$ decay contributes another 3.0\%, and
the branching fraction  
${\cal{B}}(\Lambda \rightarrow p\pi^-)$ has an uncertainty of 0.8\% 
\cite{PDG}.
The total uncertainty from the tracking reconstruction efficiency
is  4.2\%. 
Other sources of the systematic uncertainties
are:
the trigger efficiency (0.5\%), 
selection criteria (4.0\%), 
MC statistics (0.7\%) and luminosity (1.4\%). 
The trigger efficiency is estimated 
using the trigger simulation.
Assuming no correlation between them,
all these uncertainties 
are combined in quadrature to 
give a total of $9.1\%$ {in the detection sensitivity}.  
An additional systematic uncertainty due to the estimation of the
expected background
includes those due to its statistical error and the background shape.
By varying the assumptions about the background shape for the MC 
events, 
we checked that this effect is negligible compared to the 
background statistical error.

\begin{table}
\begin{tabular}{|c|c|} \hline
$\Lambda$ reconstruction & 6.0 \\
Proton identification & 3.0 \\
Branching fraction ${\cal{B}}(\Lambda \rightarrow p\pi^-)$ & 0.8 \\
Tracking reconstruction efficiency & 4.2 \\
Selection criteria & 4.0 \\
Trigger efficiency & 0.5 \\
Luminosity & 1.4 \\
MC statistics & 0.7 \\ \hline
Total & 9.1 \\ \hline
\end{tabular}
\caption{
Sources of systematic uncertainties for the detection sensitivity 
$2\varepsilon N_{\tau\tau}{\cal B}(\Lambda\rightarrow p\pi^-)$
(in \%)
}
\label{tbl:sys}
\end{table}

While the angular distribution of $\tau^-\rightarrow \bar{\Lambda}\pi^-$
and $\tau^-\rightarrow \Lambda \pi^-$ decay is initially
assumed to be uniform {in this analysis},
it is sensitive to the lepton flavor violating interaction
structure~\cite{LFV}, 
similar to other analyses~\cite{othertau}.
The spin correlation 
{between the $\tau$ lepton in the signal and that in tag side}
must be considered.
A possible nonuniformity was taken into account by comparing
the uniform case with those assuming $V-A$ and $V+A$ interactions,
which result in the maximum possible variations.
No statistically significant difference in 
the $M_{\rm inv}$ -- $\Delta{E}$
distribution or the efficiencies is found compared 
the case of the uniform distribution.
Therefore,
systematic uncertainties due to these effects 
are neglected in {the} upper limit evaluation.


The upper limits on the branching fractions at the 90\% C.L.
taking into account 
systematic uncertainties on both expected background 
and the detection sensitivity 
are then calculated using the
POLE program~\cite{cite:pole}, which integrates over Gaussian systematic errors,  
to be:
\begin{eqnarray*}
&&{\cal B}(\tau^-\rightarrow \bar{\Lambda}\pi^-) < 1.4 \times 10^{-7} \\
&&{\cal B}(\tau^-\rightarrow \Lambda \pi^-) < 0.72 \times 10^{-7}.
\end{eqnarray*}

\section{Conclusion}

In conclusion,
we have searched for 
the decay modes that violate
both lepton and baryon number 
conservation:
$\tau^-\rightarrow \bar{\Lambda}\pi^-$
($B-L$ conserving)
and 
$\tau^-\rightarrow \Lambda \pi^-$ 
($B-L$ violating)
using data collected 
by the Belle detector at the KEKB $e^+e^-$ asymmetric-energy collider.
We found no signal in the either mode.
The following  upper limits on
the branching fractions were obtained:
${\cal{B}}(\tau^-\rightarrow \bar{\Lambda}\pi^-) < 1.4\times 10^{-7}$ 
and 
${\cal{B}}(\tau^-\rightarrow \Lambda \pi^-) < 0.72\times 10^{-7}$ 
at the 90\% confidence level. 
These are the first results ever reported for these modes.

\section*{Acknowledgments}
We thank the KEKB group for the excellent operation of the
accelerator, the KEK cryogenics group for the efficient
operation of the solenoid, and the KEK computer group and
the National Institute of Informatics for valuable computing
and Super-SINET network support. We acknowledge support from
the Ministry of Education, Culture, Sports, Science, and
Technology of Japan and the Japan Society for the Promotion
of Science; the Australian Research Council and the
Australian Department of Education, Science and Training;
the National Science Foundation of China under contract
No.~10175071; the Department of Science and Technology of
India; the BK21 program of the Ministry of Education of
Korea, and the CHEP SRC program and Basic Reserch program 
(grant No. R01-2005-000-10089-0) of the Korea Science and
Engineering Foundation; the Polish State Committee for
Scientific Research under contract No.~2P03B 01324; the
Ministry of Science and Technology of the Russian
Federation; the Ministry of Higher Education, Science and Technology of the Republic of Slovenia;  the Swiss National Science Foundation; the National Science Council and
the Ministry of Education of Taiwan; and the U.S.\
Department of Energy.


\begin{thebibliography}{99}

\bibitem{cite:susy}
S.~Lola and G.~G.~Ross,
Phys.\ Lett.\ {\bf B 314}, 336 (1993).

\bibitem{cite:string}
G.~Lazarides, C.~Panagiotakopoulos and Q.~Shafi,
Nucl.\ Phys.\  {\bf B 278}, 657 (1986).

\bibitem{PDG}
S.~Eidelman  {\it et al.} (Particle Data Group),
Phys. Lett.  {\bf B 592}, 1 (2004).

\bibitem{cite:tau02}
W.~Marciano {\it et al.},
Nucl. Phys. (Proc. Suppl.) {\bf B 40}, 3 (1995).


\bibitem{protontau}
R.~Godang {\it et al.}  (CLEO Collaboration),
Phys.\ Rev.\  {\bf D 59}, 091303 (1999).

\bibitem{cite:Hou:2004uc}
W.~S.~Hou, M.~Nagashima and A.~Soddu,
arXiv:hep-ph/0509006.



\bibitem{kekb}
S.~Kurokawa and  E.~Kikutani, Nucl. Instr. Meth.  {\bf A 499}, 1
(2003), and other papers included in this Volume.
  


\bibitem{Belle}
A.~Abashian {\it et al.} (Belle Collaboration),
Nucl. Instr. and Meth. {\bf A 479}, 117 (2002).


\bibitem{EID} 
        K. Hanagaki {\it et al.},
	Nucl. Instr. and Meth.  {\bf A 485}, 490 (2002).



\bibitem{MUID} 
{A. Abashian {\it et al.}},
	Nucl. Instr. and Meth. {\bf A 491}, 69 (2002).



\bibitem{cite:koralb_tauola}
S.~Jadach and Z.~W\c{a}s, 
Comp. Phys. Commun. {\bf 85}, 453 (1995).

\bibitem{cite:qq}
QQ is an event generator developed by the CLEO Collaboration and
described in {\sf{http://www.lns.cornell.edu/public/CLEO/soft/QQ/}}.
It is based on the LUND Monte Carlo for jet fragmentation and 
$e^+e^-$ physics described in
T.~Sj$\ddot{\mbox{o}}$strand, Comp. Phys. Commun. {\bf 39}, 347 (1986)
and
T.~Sj$\ddot{\mbox{o}}$strand, Comp. Phys. Commun. {\bf 43}, 367 (1987).

\bibitem{BHLUMI} 
        S.~Jadach {\it et al.},
	Comp. Phys. Commun. {\bf 79}, 305 (1992).

\bibitem{KKMC} 
        S.~Jadach {\it et al.},
	Comp. Phys. Commun. {\bf 130}, 260 (2000).


\bibitem{AAFH} 
        F.~A.~Berends {\it et al.},
	Comp. Phys. Commun. {\bf 40}, 285 (1986).


\bibitem{cite:geant3}
R. Brun et al.,
GEANT 3.21 CERN Report No. DD/EE/84-1, 453.


%
%

\bibitem{thrust}
S.Brandt {\it et al.},
	Phys.\ Lett.\ {\bf 12}, 57 (1964);
E.Farhi,
	Phys.\ Rev.\ Lett.\ {\bf 39}, 1587 (1977).

\bibitem{Lambda_rec}
        K. Abe {\it et al.}  (Belle Collaboration),
	Phys.\ Rev.\ {\bf D 65}, 091103 (2002).

\bibitem{cite:FC}
	G.J. Feldman and R.D. Cousins, 
	Phys.\ Rev.\  {\bf D 57}, 3873 (1998).


\bibitem{LFV}
        R.~Kitano and T.~Okada,
	Phys.\ Rev.\  {\bf D 63}, 3873 (2001).


\bibitem{othertau}
For example,
K. Inami {\it et al.}  (Belle Collaboration),
Phys.\ Rev.\ Lett. \  {\bf 92}, 171802 (2004);
Y. Enari {\it et al.}  (Belle Collaboration),
Phys.\ Rev.\ Lett. \  {\bf 93}, 081803 (2004).



\bibitem{cite:pole}
	J. Conrad {\it et al.},
	Phys.\ Rev.\ {\bf D 67}, 012002 (2003).

\end{thebibliography}
\end{document}